\documentclass[pra,aps,twocolumn,notitlepage,showpacs,amsmath,amssymb]{revtex4-1}

\usepackage{graphicx}% Include figure files
\usepackage{dcolumn}% Align table columns on decimal point
\usepackage{bm}% bold math
\usepackage{array} % for better arrays (eg matrices) in maths
\usepackage{url}

\usepackage{subfigure}% Multipart figures
\usepackage{pdfsync} 

\begin{document}
\preprint{DRAFT VERSION 1.1}

\title{Effects of magnetic dipole-dipole interactions in atomic Bose-Einstein condensates with tunable s-wave interactions}% Force line breaks with \\

\author{Abraham J. Olson}
 \email{olsonaj@purdue.edu}
\author{Daniel L. Whitenack}
% \email{dwhitena@purdue.edu}
\author{Yong P. Chen}%
% \email{yongchen@purdue.edu}
\affiliation{%
Department of Physics, \\Purdue University, West Lafayette IN 47907
}%

\date{\today}% It is always \today, today,

\begin{abstract}
The s-wave interaction is usually the dominant form of interactions in atomic Bose-Einstein condensates (BECs). Recently, Feshbach resonances have been employed to reduce the strength of the s-wave interaction in many atomic speicies. This opens the possibilities to study magnetic dipole-dipole interactions (MDDI) in BECs, where the novel physics resulting from long-range and anisotropic dipolar interactions can be explored. Using a variational method, we study the effect of MDDI on the statics and dynamics of atomic BECs with tunable s-wave interactions. We benchmark our calculation against previously observed MDDI effects in $^{52}$Cr with excellent agreement, and predict new effects that should be promising to observe experimentally. A parameter of magnetic Feshbach resonances, $\epsilon_{dd,\text{max}}$, is used to quantitatively indicate the feasibility of experimentally observing MDDI effects in different atomic species. We find that strong MDDI effects should be observable in both in-trap and time-of-flight behaviors for the alkali BECs of $^{7}$Li, $^{39}$K, and $^{133}$Cs. Our results provide a helpful guide for experimentalists to realize and study atomic dipolar quantum gases. 

\end{abstract}

\maketitle

\section{Introduction}
The physics of ultracold dipolar quantum gases is a rich and promising area of research. There is great interest in the physical behaviors that result from dipole-dipole interactions \cite{Lahaye_2009_arXiv,Baranov_2008_PhysRep}. An ultracold atomic gas of atoms that posses a magnetic moment will have a magnetic dipole-dipole interaction (MDDI). However, it is often difficult to observe the effects of MDDI in many atomic species (particularly alkalis), where the MDDI is typically much weaker than the isotropic, s-wave interactions. For atomic species with magnetic Feshbach resonances, however, the s-wave scattering length, $a_s$, can be tuned via magnetic fields \cite{Kohler_2006_RMP}. Employing Feshbach resonances has led to fruitful and impressive developments in ultracold atom research. Of interest here is that it allows for the exploration of MDDI in Bose-Einstein condensates (BECs). By tuning $a_s$ to near zero, MDDI can become the strongest interaction in the BEC. 

The effect of MDDI in ultracold atomic gases was first observed using a BEC of $^{52}$Cr atoms, which posses a strong magnetic moment, $\mu$, of 6 Bohr magnetons ($\mu_B$) \cite{Stuhler_2005_PRL}. The MDDI were observed to affect the aspect ratio of the $^{52}$Cr BEC in time-of-flight (TOF) free expansion. The MDDI effect on the stability of a BEC was also experimentally studied in $^{52}$Cr, where for various trap configurations the MDDI either made the BEC more or less stable against collapse \cite{Koch_2008_NatPhys}. More recently, in $^{7}$Li the MDDI effect was seen when comparing the axial length of the BEC near $a_s = 0$ for two different trapping geometries \cite{Pollack_2009_PRL}. Effects of MDDI have also been observed on the decoherence rate in a $^{39}$K BEC atomic interferometer \cite{Fattori_2008_PRL}, and on the spin domains of a spinor $^{87}$Rb Bose gas \cite{Vengalattore_2008_PRL}. Very recently, a BEC has been realized with $^{164}$Dy, with a strong magnetic moment of 10$\mu_B$, exhibiting dipolar effects even with no tuning of the scattering length \cite{Lu_PRL_2011}. Many further effects due to MDDI have been predicted such as the excitation of collective modes by tuning the dipolar interaction \cite{Giovanazzi_2002_PRL}, the emergence of a biconcave structure with local collapse \cite{Ronen_2007_PRL} \cite{Wilson_2009_arXiv}, and the modification of the phase diagram of dipolar spin-1 BECs \cite{Yi_2004_PRL} \cite{Yi_2006_PRL} \cite{Yi_2006_PRA},  of the soliton stability in 1D BECs \cite{Gligoric_2009_PRL}, and of vortices in BECs \cite{Abad_2009_PRA}. We refer the reader to recent reviews for a further discussion of the multiplicity of MDDI effects \cite{Baranov_2008_PhysRep, Lahaye_2009_arXiv}.

Motivated by such rich physics, we theoretically model the effects of MDDI and possibility of experimentally detecting such effects in BECs of $^{52}$Cr and all the alkalis. In Section II, we explain the variational method we employ to model MDDI using a cylindrically-symmetric, Gaussian Ansatz for the BEC wave-function \cite{Yi_2001_PRA}. In Section III,  we  start by discussing the relevant parameters for our simulations. We also introduce a key quantity used in this paper, the ratio of s-wave scattering length, $a_s$, to a length  defined for MDDI, $a_{dd}$. Next in that section we present simulations of the effects of MDDI in $^{52}$Cr, including those reported in Refs.~\cite{Koch_2008_NatPhys} and \cite{Lahaye_2007_Nature}, and predict several additional effects that should be readily observable. We also present simulations for the effects of MDDI in $^{7}$Li, $^{39}$K and $^{133}$Cs, the three alkali BECs we identified as promising for observing MDDI effects. In Section IV, we conclude.

\section{Variational Method}
While a few other methods exist to model dipole-dipole interaction effects in BECs \cite{Santos_2000_PRL, Bao_2010_arXiv, VanBijnen_arXiv_2010,Sapina_PRA_2010,Lahaye_2009_arXiv}, the variational method we employ has shown great utility because its simple, analytic solutions are valid over a wide range of experimental parameters \cite{Garcia_PRL_1996, Yi_2001_PRA, Yi_2003_PRA}. Two types of interactions are considered. The s-wave interaction is characterized by the s-wave scattering length, $a_s$, and for dipole-dipole interactions a parameter defined as $a_{dd}$ is used. MDDI can be characterized by $a_{dd} = \mu_0 \mu^2 M / (12 \pi \hbar^2)$, where $\mu$ is the magnetic moment of the atom, $M$ its mass, and $\mu_0$ is the permeability of free space~\footnote{Electronic dipole-dipole interactions could be similarly characterized but are not considered further in this paper.}. The atom-atom interaction potential thus has two terms, one for each interaction:
\begin{equation}
	V_{atom-atom}(\vec{R}) = a_s\frac{4 \pi \hbar^2}{M} \delta (\vec{R}) + a_{dd} \frac{3 \hbar^2}{M}\frac{1-3cos^2 \theta}{R^3} 
\end{equation}
Using that interaction term, the Gross-Pitaevskii equation for a BEC takes the form (in dimensionless units): 
\begin{equation}
	\begin{split}
i \frac{\partial\psi (\vec{r})}{\partial t} = -\frac{1}{2}\nabla^2 \psi(\vec{r}) + V_{ext}(\vec{r}) \psi(\vec{r}) + \frac{4 \pi N a_s}{a_{r}} |\psi(\vec{r})|^2 \psi(\vec{r}) \\ +\frac{N a_{dd}}{3 a_{r}} \int \frac{1-3 cos^2 \theta}{R^3} |\psi(\vec{r'})|^2 d\vec{r'}  \psi(\vec{r})
\end{split}
\end{equation}
where the length unit is $a_r =  \sqrt{\hbar / m\omega_r}$, the trap frequencies are $\omega_r$ and $\omega_z$ for the respective radial and axial direction, the time unit $ t=2\pi/\omega_r$, $\psi$ is normalized to unity ($|\psi|^2 = 1$), and $V_{ext}(\vec{r}) =   \frac{1}{2} (x^2 + y^2 + \lambda z^2)/2 $ is the trap potential where the trap aspect ratio is $ \lambda =  \omega_z / \omega_r$. For this paper, we adopt the nomenclature for trap shapes that is common to experiments: the symmetrical a.k.a. spherical ($\lambda = 1$), the cigar a.k.a. prolate ($\lambda < 1$), and the pancake a.k.a. oblate ($\lambda > 1$) shapes \cite{Koch_2008_NatPhys}; we also assume that the magnetic field is applied along the axial direction. Using a cylindrical-symmetric, Gaussian Ansatz for the BEC wave-function \footnote{We note that there are limitations to the Gaussian anzatz. For example, it does not account for biconcave structure predicted to occur in pancake traps. However, this is an issue more specific to the pancake traps and only in a certain region of interactions\cite{Ronen_2007_PRL}. For most trap geometries (especially cigar traps) and properties studied in this paper, the Gaussian ansatz provides a good model to guide the experimental investigation of MDDI effects.}
\begin{equation}
	\psi \left(\vec{r}\right)= \exp\left(-\frac{x^2}{2 q_r^2}\right) \exp\left(-\frac{y^2}{2 q_r^2}\right) \exp\left(-\frac{z^2}{2 q_z^2}\right)
\end{equation}
the variational method results in two differential equations that describe the mean axial, $q_z$, and radial, $q_r$, lengths of the BEC (detailed solution in Ref.~\cite{Yi_2001_PRA}, noting \footnote{Typo in \cite{Yi_2001_PRA, Yi_2003_PRA} corrected. (S. Yi,  Private Communications)}):
\begin{subequations}
\label{eqn:diffeqs}
\begin{eqnarray}
\ddot{q_r} + q_r & = & \frac{1}{q^3_r} - \sqrt{\frac{2}{\pi}}\frac{1}{q_r^3 q_z} \frac{N}{a_{\rho}} [a_{dd} f(\kappa) - a_s]\label{eqn:diffeqsRadial} \\ 
\ddot{q_z} +\lambda^2 q_z & = & \frac{1}{q_z^3} - \sqrt{\frac{2}{\pi}}\frac{1}{q_r^2 q_z^2}\frac{N}{a_{r}} [a_{dd} g(\kappa) - a_s] \label{eqn:diffeqsAxial}
 \end{eqnarray}
 \end{subequations}
where
\begin{subequations}
\label{eqn:diffeqsSubEqns}
\begin{eqnarray}
q_r & \equiv & \sqrt{\frac{<x^2>}{2}} \equiv \sqrt{\frac{<y^2>}{2}} \\
q_z &\equiv & \sqrt{\frac{<z^2>}{2}} \\
\kappa & \equiv & q_r / q_z  \mbox{, the BEC aspect ratio} \\
f(\kappa) &\equiv & \frac{[-4 \kappa^4-7\kappa^2+2+9\kappa^4 H(\kappa)]}{2(\kappa^2-1)^2} \\
g(\kappa) &\equiv & \frac{[-2\kappa^4+10\kappa^2 + 1-9\kappa^2 H(\kappa)]}{(\kappa^2-1)^2} \\
H(\kappa) &\equiv & \frac{\tanh^{-1} \left( \sqrt{1-\kappa^2}\right)}{ \sqrt{1-\kappa^2}}
\end{eqnarray}
\end{subequations}

These coupled differential equations model BEC behavior with (keeping $a_{dd}$) and without (setting $a_{dd}=0$) MDDI effects, and they can be numerically solved to model three experimentally relevant situations: 
\begin{enumerate}
	\item The static, in-situ sizes for a trapped BEC are found by setting the time-dependent components, $\ddot{q_r}$ and $\ddot{q_z}$, to zero. 
	\item In-trap dynamics are modeled by keeping all terms.
	\item Time-of-flight (TOF) free expansion behavior is modeled by removing the the terms $q_r$ and $\lambda^2 q_z$, which represent the trapping potential, on the left sides of Eqns.~\ref{eqn:diffeqs}.
\end{enumerate}
\section{Results}
In this section, we employ the above method to solve for both the in-trap and TOF behaviors of BECs with MDDI \footnote{Code developed for this were written in \texttt{Matlab}, and are available at \protect\url{www.physics.purdue.edu/quantum/MDDI.html} and \protect\url{www.abeolson.com/physics/MDDI.html} }. We also find the threshold $a_s$, denoted $a_s^{\mathrm{threshold}}$, below which the BEC is unstable and collapses. We benchmark our variational calculations against available experimental results in $^{52}$Cr and find good agreement (Section III.B). We also make predictions of MDDI effects in the alkalis, and find that $^7$Li, $^{39}$K, and $^{133}$Cs are the species most favorable for the exploration of MDDI effects. 
\subsection{Parameters}
The input parameters used in our simulation are the number of atoms in the trap ($N$), the magnetic dipole moment of the atom ($\mu$), the mass of the atom ($M$), the axial ($f_z$) and radial ($f_r$) frequencies of the trap ($2 \pi f_{r,z} = \omega_{r,z}$), and the s-wave scattering length ($a_s$). An applied magnetic field can tune $a_s$ via a Feshbach resonance with an analytic approximation given by $a_{s} (B)=a_{bg}\left(1-\frac{\Delta}{B-B_{\infty}}\right)$, where $\Delta$ is the width and $B_{\infty}$ is the location of the Feshbach resonance,  and $a_{bg}$ is the scattering length far from any resonances. An experimental limit for reaching small $a_s$ is the precision of control over the magnetic fields. In typical ultracold atom experiments an experimental precision $\delta B/B$ of approximately $10^{-5}$ can be realized \cite{Marte_2002_PRL, Chin_2000_PRL, DeMarco_2001_thesis}. As $\mu$ depends on the strength of the magnetic field, we calculate $\mu$ at $B_0=B_{\infty}+\Delta$ (where $a_s=0$), denoted $\mu_{cross}$. For reference, some parameters for known Feshbach resonances from the literature are listed along with the calculated values of $\mu_{cross}$ in Appendix~\ref{append:experimentalParameters} Table~\ref{tab:magneticmoment}.

To compare the potential for experimentally observing MDDI effects in the various atomic species of interest, we employ the dimensionless ratio $\epsilon_{dd}=a_{dd}/a_s$ used in Ref. \cite{Lahaye_2007_Nature}, focusing on its maximal value that can be achieved experimentally:
\begin{equation}\label{eqn:epsilondd}
	\epsilon_{dd\text{, max}} \equiv \frac{a_{dd}}{a_{s,\text{min}}}=\frac{\mu_0 \mu^2 m}{12 \pi \hbar^2 a_{bg}(\delta B / \Delta)}
\end{equation} 
where $a_{s,\text{min}}\approx a_{bg} \left(\delta B / \Delta \right)$ is the minimal $a_s$ that can be achieved given a typical experimental magnetic field stability (assumed to be $\delta B/B_0 \approx 10^{-5}$).
The alkali species that are the best candidates for observing and studying MDDI effects in BECs are clearly $^7$Li and $^{39}$K and $^{133}$Cs (see Table~\ref{tab:epsilondd}). We note that $^{52}$Cr, while having a much larger $a_{dd}$, does not have a broad Feshbach resonance that allows for as precise tuning of $a_s$ as in some of the alkalis. 
\begin{figure}[htbp]
  \includegraphics[width=0.4\textwidth]{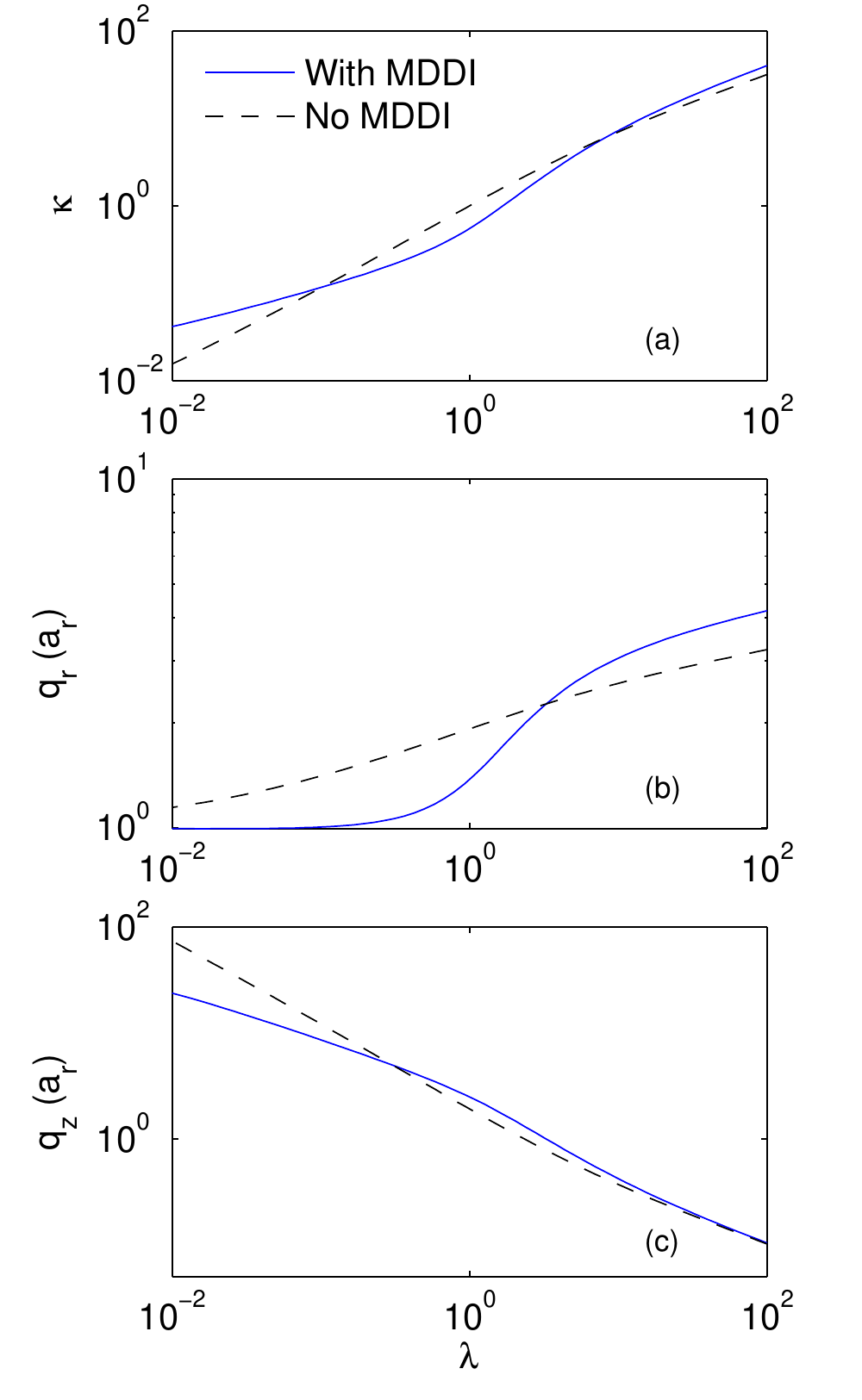}
  \caption{The calculated in-trap BEC aspect ratio \textbf{(a)}, radial length \textbf{(b)}, and axial length \textbf{(c)} of a $^{52}$Cr BEC with (solid, blue) and without (dashed, black) MDDI for a range of trap aspect ratios, $\lambda$. In this simulation, the $f_{avg} = 700$ Hz, the atom number $N= 2\times10^4$, and $a_s=15 a_0$. The magnetic field is, as for all simulations in this paper, aligned along the axial direction.}
	\label{fig:mddiInTrapARE}
\end{figure}

\begingroup
\squeezetable
\begin{table}
\begin{tabular}{ccccccc}
\hline
\hline
Species  & $|F, m_F>$ & $\mu/\mu_B\footnote{Calculated at $B_0$}$ & $a_{s, \text{min}}$ ($a_0$) & $a_{dd}$ ($a_0$) & $\epsilon_{dd, max}$ \\
\hline
$^{7}$Li & $|1,+1>$ &0.94 & 0.0007 & 0.041 & 58.4 \\
$^{23}$Na & $|1,+1>$ & 0.91 & 0.572 & 0.047 & 0.08\\
$^{39}$K & $|1,+1>$ & 0.95 & 0.0022 & 0.287 & 130 \\
$^{41}$K & $|1,-1>$ & 0.07& 0.104 & 0.002 & .020 \\
$^{85}$Rb & $|2,-2>$ & -0.57& 0.067 & 0.223 &3.33 \\
$^{87}$Rb & $|1,+1>$ & 0.73 & 5.98 & 0.374 & 0.062 \\
$^{133}$Cs & $|3,-3>$ & -0.75& 0.0096 & 0.607 & 63.2  \\
$^{52}$Cr & $|3,-3>$ & 6 & 0.473 & 15.2 & 32.2 \\
\hline
\end{tabular}
\caption{Parameters for variational computations, with the maximum value of $\epsilon_{dd}$ as calculated from Eqn.~\ref{eqn:epsilondd}.}
\label{tab:epsilondd}
\end{table}
\endgroup

As an initial verification of our model, we compute $\kappa$, in-situ ($t=0$) and in the TOF asymptotic limit ($t\rightarrow \infty$) assuming only s-wave interactions, for both strongly interaction ($a_s \rightarrow \text{Large}$) and non-interacting ($a_s=0$) cases. The expected in-trap and TOF behaviors of a BEC in such limiting cases are (see Refs.~\cite{Castin_1996_PRL, Pethick_2008_BEC}):
\begin{enumerate}
	\item If $a_s = 0$ and $t=0$, expect $\kappa=\sqrt{\frac{\omega_a}{\omega_r}} = \lambda^{1/2}$.
	\item If $a_s = \text{Large}$, $t=0$, expect $\kappa=\frac{\omega_a}{\omega_r} = \lambda$.
	\item If $a_s = 0$, $t\rightarrow \infty$, expect $\kappa=\sqrt{\frac{\omega_r}{\omega_a}} = \lambda^{-1/2}$.
	\item If $\lambda \ll 1$ or $\lambda \gg 1$, and if $a_s = \text{Large}$ with $t\rightarrow \infty$, expect $\kappa=\frac{2}{\pi} \frac{\omega_r}{\omega_a} = 2 \lambda^{-1} /\pi$.
\end{enumerate}
Our variational calculation (performed with $\mu=0$ in these cases) does reproduce these expected values for $\kappa$.

%%%%%%%%%%%%%%%End Parameters/summary%%%%%%%%%%%%%%%
\subsection{Effects of MDDI in $^{52}$Cr BEC}
The first observation of MDDI in a BEC was with $^{52}$Cr. As $^{52}$Cr is the most studied atomic species so far for MDDI effects in BECs, we present and benchmark our calculation and results for $^{52}$Cr before discussing the alkalis. We used $^{52}$Cr to demonstrate four characteristic MDDI effects, discussed in detail below.

\subsubsection{Effect of MDDI on in-situ aspect ratio of a BEC}
A calculated result of a $^{52}$Cr BEC trapped in a harmonic trap is shown in Fig.~\ref{fig:mddiInTrapARE}. For a nearly symmetrical trap ($\lambda \approx 1$, which is the case for the experiment in Ref.~\cite{Koch_2008_NatPhys}) the MDDI increase the axial length of the BEC and reduce its radial length compared to a BEC with no MDDI, leading to a decreased aspect ratio. However, if the trap is very prolate, $\lambda \ll 1$ (very oblate, $\lambda \gg 1$), we find that the BEC will shrink (expand) in both the axial and radial directions, in a way that leads to an \emph{increased} aspect ratio. There are two values of $\lambda$, one for when the trap is oblate and one for when the trap is prolate, where the MDDI do not change the aspect ratio of the BEC. Similar in-situ effects of MDDI on $\kappa$ were found in all simulations of stable BECs for the alkali as well.

\begin{figure}[htbp] \includegraphics[width=.45\textwidth]{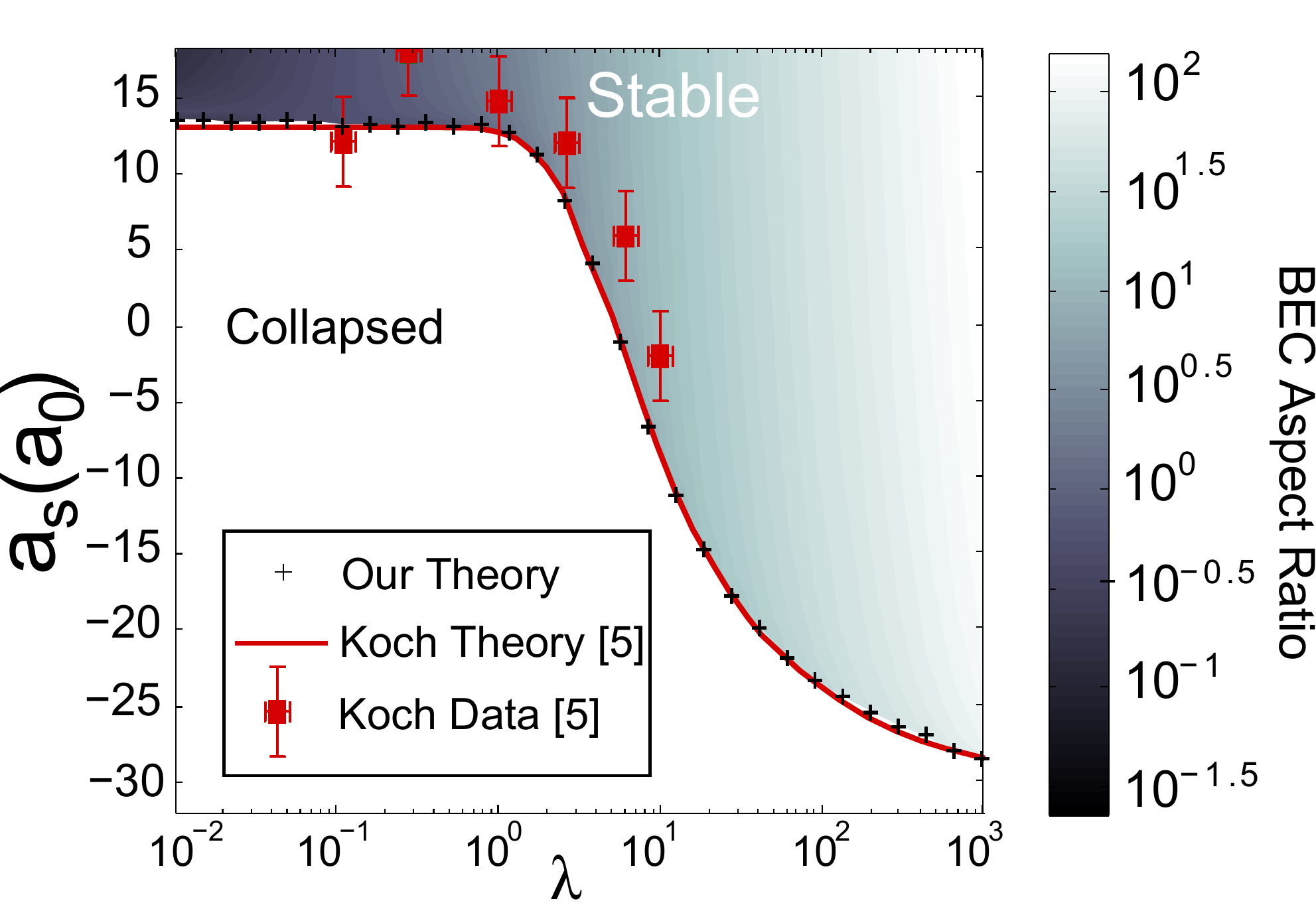}
  \caption{Stability diagram and aspect ratio of a  $^{52}$Cr BEC, with parameters ($f_{avg}= 700$ Hz, and $N=2\times10^4$) chosen to resemble those in the experiment Ref.~\cite{Koch_2008_NatPhys} . The BEC aspect ratio, $\kappa$, solved using our method, is plotted in the color map as functions of $a_s$ and $\lambda$ in the stable regime (log 10 scale). The energy variational solution and experimental data from Ref.~\cite{Koch_2008_NatPhys} for $a_s^{\mathrm{threshold}}$ are included for comparison. This shows the effectiveness of our method in solving not just the $a_s^{\mathrm{threshold}}$ where the BEC collapses but also the BEC size in the stable regime.}
	\label{fig:cr52stability}
\end{figure}

\begin{figure*}[htbp]  \includegraphics[width=0.75\textwidth]{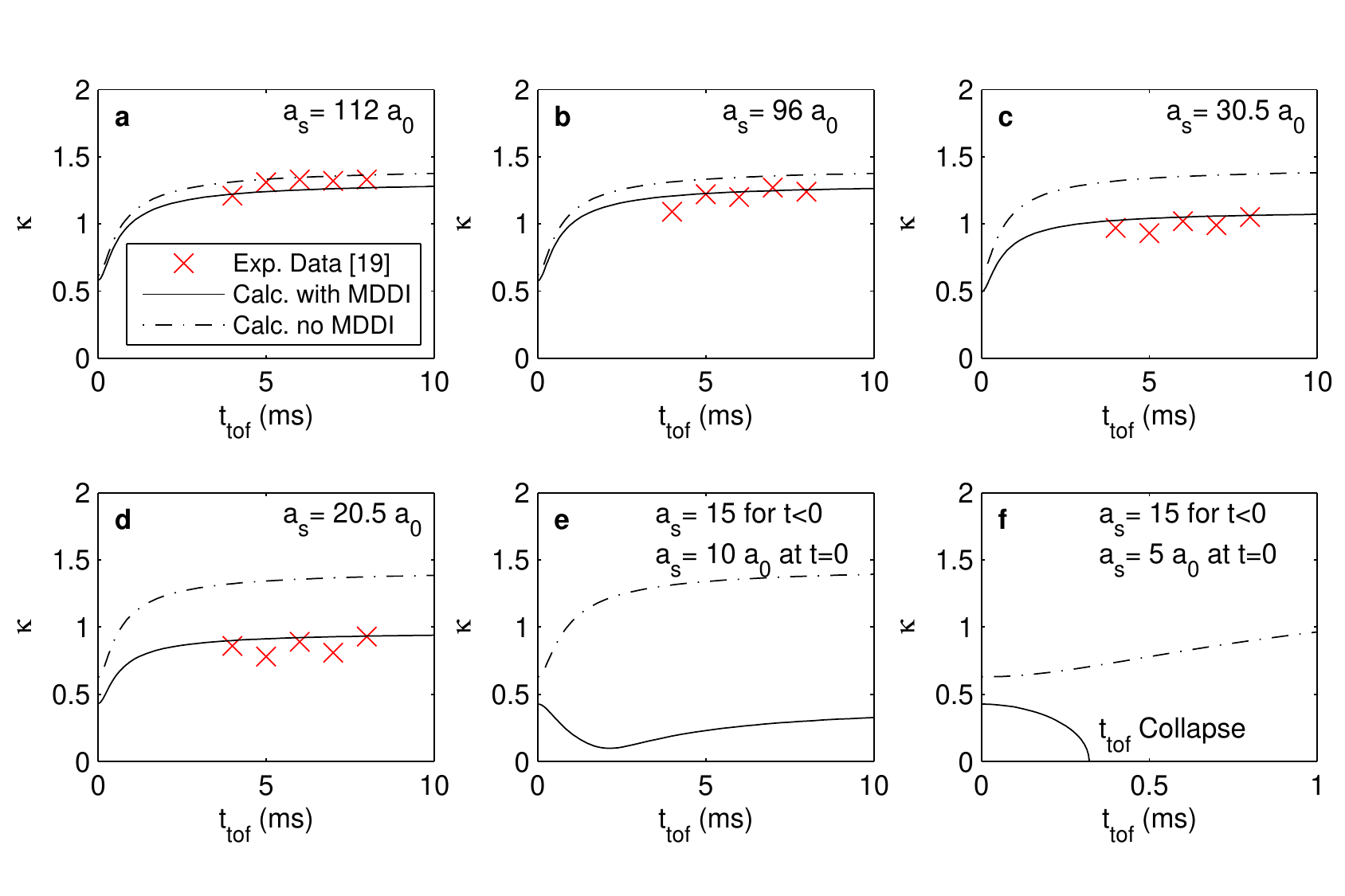}
  \caption{The aspect ratio of a $^{52}$Cr BEC vs the time-of-flight duration,$t_{TOF}$, for different scattering lengths (\textbf{a-d}) with parameters chosen to resemble those in Ref.~\cite{Lahaye_2007_Nature}. We additionally show simulated TOF behavior where MDDI nearly \textbf{(e)} and actually \textbf{(f)} causes BEC collapse. Upon the release of the BEC, the scattering length is tuned to $a_s =10 \, a_0$ in \textbf{(e)}, and $a_s =5 \, a_0$ in \textbf{(f)}. For \textbf{(a-f)}, $N=3\times10^4$, $f_r = 600$ Hz, and $f_z = 370$ Hz.}
% (96 $a_0$, 30.5$a_0$, and 20.5$a_0$) \textbf{(a)} is $\epsilon_{dd} \approx 0$, and \textbf{(b)}, \textbf{(c)}, \textbf{(d)} are $\epsilon_{dd}=0.16, 0.5, 0.75$ respectively.
	\label{fig:cr52_variousscattering}
\end{figure*}

\subsubsection{Effect of MDDI on stability of a trapped BEC}
For BECs with only s-wave interactions, it is shown that if $a_s<0$ (attractive interactions) the BEC will collapse if $N\gtrsim 0.55 a_{ho} / |a_s|$, where $a_{ho} = \sqrt{\hbar / m\bar{\omega}}$ and $\bar{\omega}$ is the average trap frequency \cite{Sackett_1998_PRL,Donley_2001_Nature,Pethick_2008_BEC}. For any purely s-wave interacting BECs held in a three dimensional harmonic trap, the BEC will not collapse if $a_s>0$. However, MDDI can destabilize or stabilize a BEC that would otherwise be stable or unstable. The effect of MDDI on the stability of a trapped BEC was first observed in $^{52}$Cr \cite{Koch_2008_NatPhys}. For cigar traps with the B-field aligned along the axial direction, MDDI can lead to BEC collapse at a larger value of $a_s$ (in some cases, even $a_s>0$) than in an otherwise identical BEC with no MDDI. For pancake traps, the effect is opposite, and the MDDI stabilize the BEC. Koch \emph{et al.} performed an experimental study of this MDDI effect~\cite{Koch_2008_NatPhys}. To model their results, Koch \emph{et al.} employed an energy argument based on a variational Gaussian Ansatz for the BEC density distribution. Their method relied on finding when a minimum in the energy vanishes as $a_s$ is reduced. Different from their method, our variational calculation solves Eqns.~\ref{eqn:diffeqsRadial} and~\ref{eqn:diffeqsAxial} over a range of $a_s$, and find the threshold $a_s$  when those equations do not have stable numerical solutions, which indicates BEC collapse. Our method allows not only the determination of the threshold $a_s$ for collapse (which agrees with those obtained in Ref.~\cite{Koch_2008_NatPhys}) but also the axial and radial BEC lengths over the entire range of $a_s$ where the BEC is stable. We employed our method, using $N$ and $f_{avg}$ identical to those used in Fig. 3 of Ref.~\cite{Koch_2008_NatPhys}, to solve for the threshold $a_s$  and also $\kappa$ for $a_s>a_s^{\mathrm{threshold}}$ over a range of $\lambda$ (see Fig.~\ref{fig:cr52stability}). We find excellent agreement with both Koch \emph{et al}'s calculation and their experimental data. Our calculation shows, as observed in the original experiment \cite{Koch_2008_NatPhys}, that the anisotropic dipole-dipole interactions cause the threshold $a_s$ to depend strongly on the trap geometry. A similar dependence of $a_s^{\mathrm{threshold}}$ on $\lambda$ is also seen later in this paper for alkali BECs.

\begin{figure}[htbp]
\includegraphics[width=.4\textwidth]{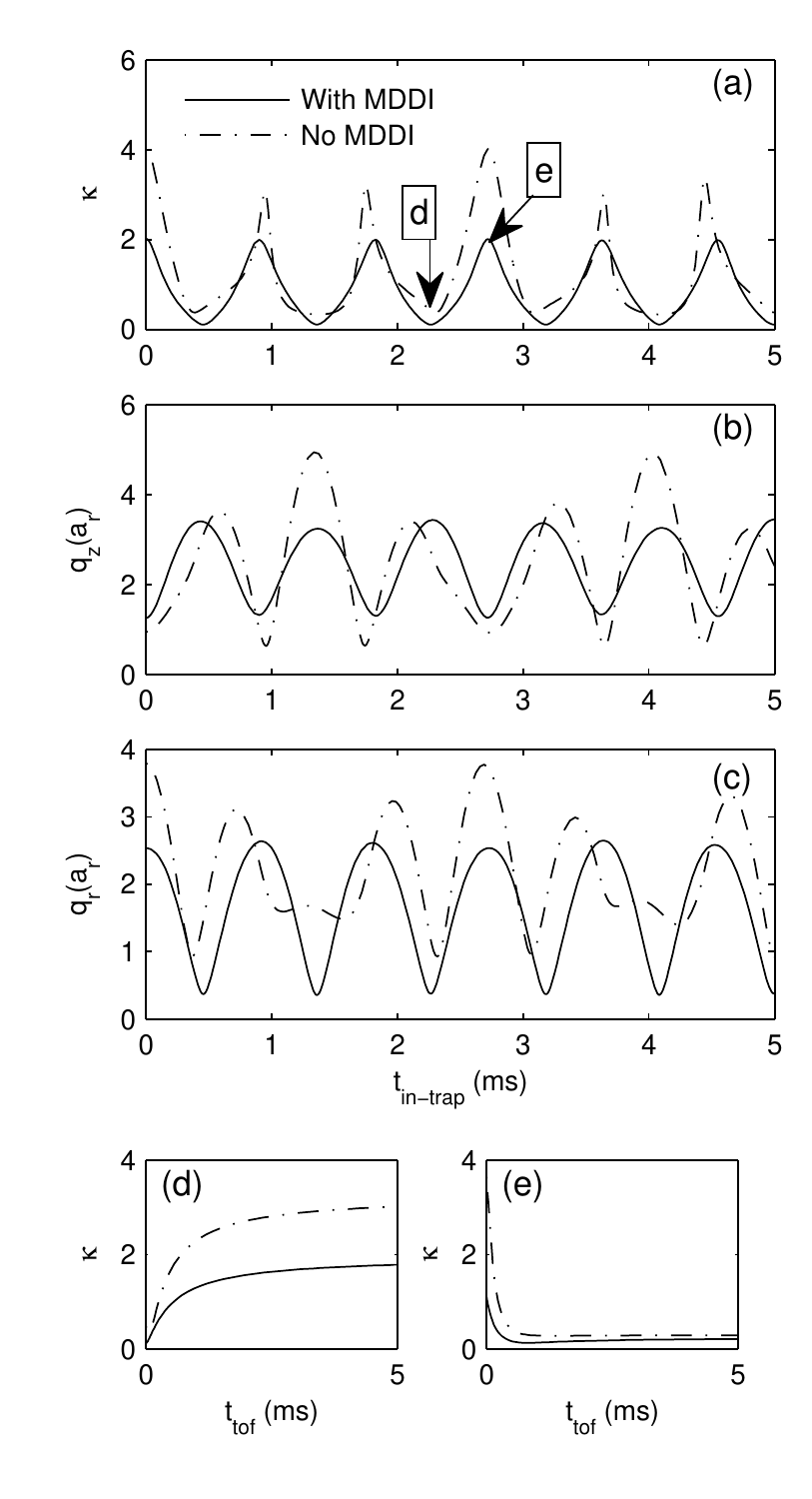}
  \caption{The aspect ratio,\textbf{(a)}, the axial length, \textbf{(b)}, and the radial length,\textbf{(c)}, for a $^{52}$Cr BEC evolving in trap with an initial radial length  twice the static value and the initial axial length half the static value. \textbf{(d)} and \textbf{(e)} show the free expansion for the BEC if it were released at times from the trap when the aspect ratio is at a minimum and maximum values respectively. The release points are indicated with arrows in \textbf{(a)}. The parameters used in \textbf{(a-e)} are $f_z=f_r=700 Hz$, $N=2\times10^4$, and $a_s=14 a_0$. }
	\label{fig:cr52evolve}
\end{figure}

\subsubsection{MDDI effect in time-of-flight behavior} 
MDDI affects the behavior of a BEC released from a trap, and such an effect was first observed in an experiment on $^{52}$Cr, where the BEC's aspect ratio in TOF changed when the applied magnetic field was along the axial verses radial direction \cite{Stuhler_2005_PRL} \cite{Lahaye_2007_Nature}.

We have simulated the experiment in Ref.~\cite{Lahaye_2007_Nature} (specifically the data shown in their Fig. 4) and find good agreement.  We assume cylindrically-symmetric ($f_r = 600$ Hz and $f_z = 370$ Hz) trap for ease of calculation with parameters approximating their ``trap 2'', which had frequencies $f_x$,$f_y$,$f_z$=660,540,370 Hz, and N$=3\times10^4$. The values of $a_s$ (112 $a_0$, 96 $a_0$, 30.5$a_0$, 20.5$a_0$) and resulting  $\epsilon_{dd}$ are chosen to match the values in Fig. 4  \textbf{a-d} of \cite{Lahaye_2007_Nature} ($\epsilon_{dd} =$ 0.14, 0.16, 0.5, 0.75). Even with the cylindrical trap approximation, the simulation results agree with the observed data quite well, and clearly shows the effect of MDDI to reduce the aspect ratio of the BEC in TOF from this trap configuration, as shown in Fig.~\ref{fig:cr52_variousscattering}. We further show that in the case of small $a_s$, MDDI-induced collapse could also be observed after a BEC is released from a trap if the $a_s$ is tuned to a small value upon release. Fig.~\ref{fig:cr52_variousscattering} also shows the MDDI-induced near-collapse \textbf{(e)} and the collapse \textbf{(f)} of atoms in free-expansion.\footnote{We have verified that typical experimental limitations in changing $a_s$ will not limit the observation of this effect, as collapse still occurs even when $a_s$ is changed after a short (a few 100's of $\mu$s for this simulation) TOF.}

\subsubsection{Effect of MDDI on in-trap dynamics}
Our model can also solve the in-trap dynamics of  BECs with MDDI. A simulation of the aspect ratio, radial size ($q_r$), and axial size ($q_a$) of a trapped $^{52}$Cr BEC---initially perturbed from its static state---evolving with time is shown in Fig.~\ref{fig:cr52evolve}, revealing an oscillatory behavior. One effect of the MDDI in this situation is to reduce amplitude of the oscillations. The oscillations may be difficult to observe in-situ due to the small condensate size, but would become easier to observe in TOF measurements taken from different instants of the oscillations. As seen in Fig.~\ref{fig:cr52evolve} \textbf{(d)} and \textbf{(e)}, the aspect ratio in TOF changes by a factor of about 2 because of the MDDI. An in-depth treatment of the modes of oscillatory, in-trap dynamics of a BEC with MDDI can be found in Refs.~\cite{ODell_PRL_2004} and \cite{Giovanazzi_PRA_2007}.

\begin{figure*}[htbp]
  \includegraphics{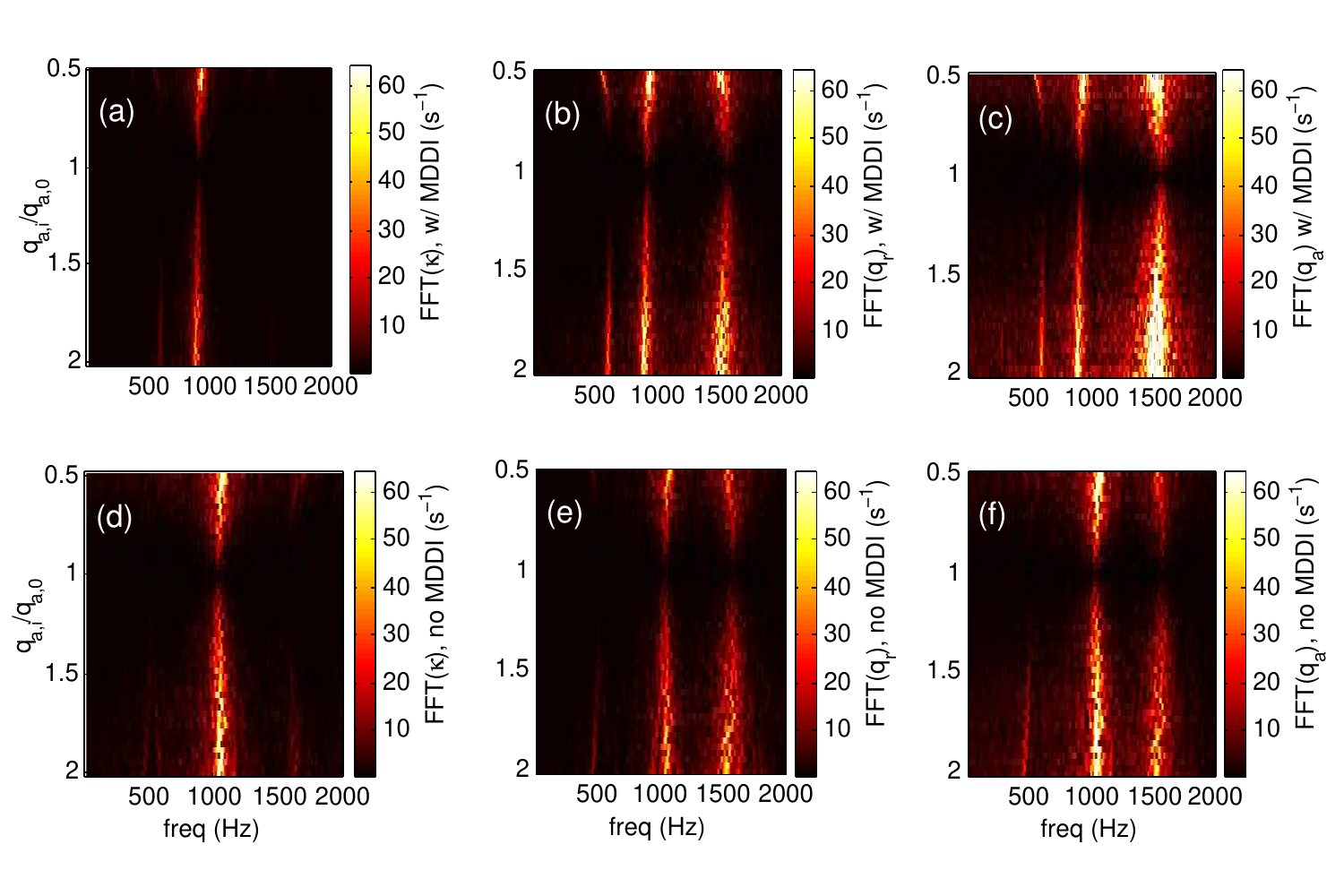}
  \caption{Simulation of a $^{52}$Cr BEC frequency spectra of the in-trap oscillations. Frequency spectra are obtained by taking the Fourier transform of the in-trap oscillations of the aspect ratio $\kappa$ \textbf{(a,d)}, radial length $q_r$ \textbf{(b,e)}, and the axial length $q_z$ \textbf{(c,f)} that occur after changing the axial size to $q_{a,i}$ from its in-trap equilibrium value, $q_{a,0}$. The colorbar gives the amplitude of the FFT for each plot. Note the shifts in both amplitude and frequency of oscillations between the simulations including \textbf{(a-c)} and not including \textbf{(d-f)} MDDI. The parameters assumed a spherical trap with $f_r=f_a=700$Hz, N=$2\times 10^4$, and $a_s=14 a_0$.}
	\label{fig:cr52FFTspec}
\end{figure*}

MDDI effects have been observed in collective oscillations of $^{52}$Cr BECs \cite{Bismut_PRL_2010}. While the results of Ref.~\cite{Bismut_PRL_2010} cannot be directly simulated by our method because their magnetic field was not aligned on the axial direction of the BEC, we are motivated by this experiment to study the collective oscillations of BECs with MDDI. By taking the Fourier transform of the time-dependent, in-trap oscillations, we obtain the frequency spectra of the collective oscillations of a BEC with MDDI, exemplified in Fig~\ref{fig:cr52FFTspec}. The three lowest modes of oscillation were observed, and shifts due to MDDI were seen in both the amplitude and frequency of the modes.

%%%%%%%%%%%%%%%%%%END OF CR52 OVERVIEW. ONTO THE ALKALIS,

\subsection{Results for alkalis}

\subsubsection{Highlight of MDDI effects in Alkali BECs: stability}
\begin{figure}[hbtp]
     \includegraphics[width=0.5\textwidth]{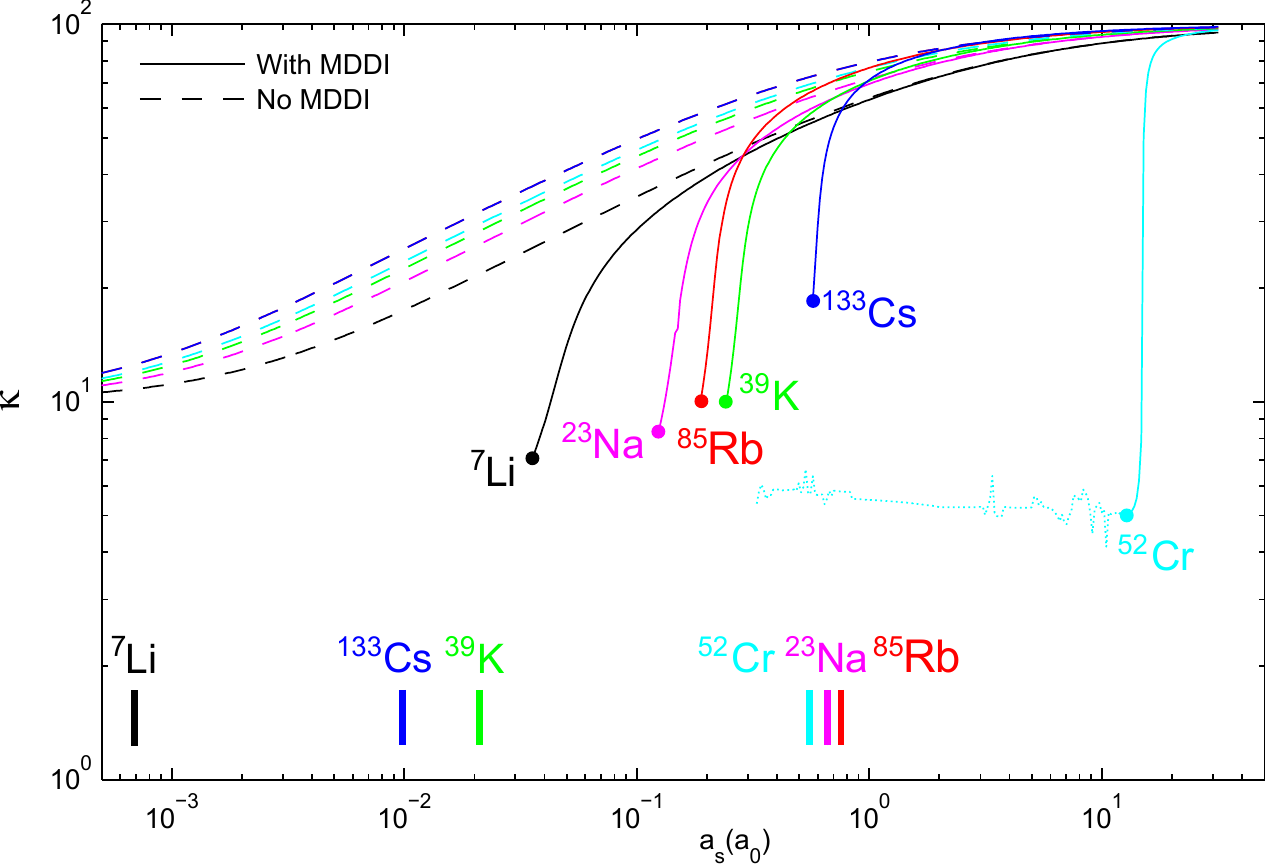}
	 \caption{The in-trap aspect ratio for a BEC near $a_s=0$. The colored, vertical bars indicate the minimum scattering length currently achievable in typical experiments, $a_{s,\text{min}}$. Calculation for each species assumes a cigar trap with $\lambda=100$ where $f_z=20$ Hz, $f_r = 2000$ Hz, and $10^6$ atoms in the BEC. A thick dot at the end of each solid curve indicates the $a_s^{\mathrm{threshold}}$, which depends both on the mass and the magnetic moment of the atom. The dotted line, shown only for $^{52}$Cr as a demonstrative example, indicates the numerically unstable solutions for $a_s < a_s^{\mathrm{threshold}}$ and is used to determine the threshold $a_s$. Collapse threshold values from the smallest to largest are (0.01, 0.02, 0.20, 0.24, 0.54, 13.50)$\times \,a_0$ for ($^7$Li, $^{23}$Na, $^{85}$Rb, $^{39}$K, $^{133}$Cs, $^{52}$Cr) respectively.} 
     \label{fig:AllYasp}
\end{figure}

A central point of our paper is that MDDI effects are also possible to observe in BECs of the alkalis, even with a much smaller $\mu$ than $^{52}$Cr. To show this, we compare dipolar collapse for BECs of various species in Fig.~\ref{fig:AllYasp}. If the MDDI are not included in the model, then the BECs are stable for any positive value of $a_s$. However, with MDDI the BEC aspect ratio decreases more substantially as $a_s$ is reduced towards zero. The BEC collapses beyond a $a_s^{\mathrm{threshold}}$, indicated by the heavy dot. The value of $a_s^{\mathrm{threshold}}$ can be compared with the $a_{s,\text{min}}$ of Table \ref{tab:epsilondd}, indicated by the colored bars in the figure. This comparison clearly indicates whether the observation of collapse due to MDDI is feasible with current experimental abilities and reveals $^{7}$Li, $^{39}$K, and $^{133}$Cs as promising alkali species for such an observation.

\subsubsection{$^{7}$Li, $^{39}$K, and $^{133}$Cs}
We find that $^{7}$Li, $^{39}$K, and $^{133}$Cs possess the greatest potential among alkali BECs for exploring MDDI effects. Several examples of such effects are shown in the simulations below. We describe $^7$Li in detail, and similar results are presented for $^{39}$K and $^{133}$Cs. The $|1,+1>$ state of $^{7}$Li is excellent for studying MDDI effects as it has the the widest known Feshbach resonance of the alkali atoms; the slope at zero-crossing is only $\approx 0.1 a_0/G$ \cite{Pollack_2009_PRL}. Using our variational method, we find that each of the MDDI effects discussed so far in $^{52}$Cr should be observable within current experimental capabilities with $^7$Li as well. 

\begin{figure}[htbp]
     \includegraphics[width=3.2in]{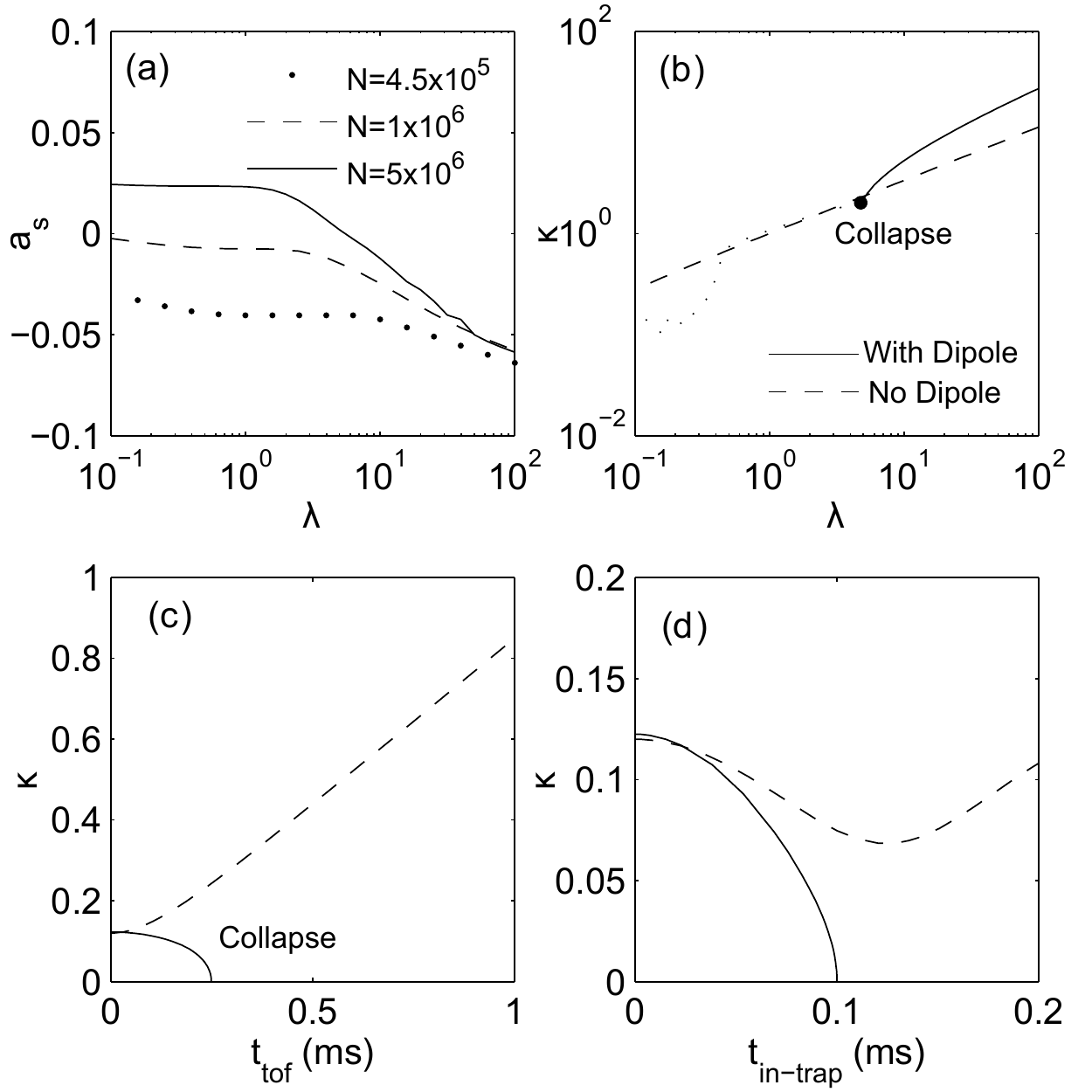}
	 \caption{Effect of MDDI in a $^7$Li BEC. \textbf{(a)} the calculated threshold $a_s$ between stable and unstable regimes for a range of $\lambda$ and three representative atom numbers. \textbf{(b)} the BEC aspect ratio over a range of $\lambda$, with the collapse regime indicated by the faint dotted line, for which there is no stable solution found for Eqns. \ref{eqn:diffeqsRadial} and \ref{eqn:diffeqsAxial}. \textbf{(c)} Evolution of the aspect ratio of the BEC in TOF expansion after release from a trap. In the trap $a_s = 0.1 a_0$ (where the BEC is stable), and upon release $a_s$ is tuned to $0.001 a_0$ to induce collapse in free expansion. \textbf{(d)} the in-trap evolution of a BEC similar to \textbf{(c)} with $a_s$ tuned to $0.001 a_0$ at $t=0$ to induce in-trap dipolar collapse. Parameters used in the simulation: \textbf{(a)} and \textbf{(b)} $f_{avg}=700$ Hz; \textbf{(b)}, $N=5\times10^6$ and $a_s = 0.001 a_0$; \textbf{(c)} and \textbf{(d)}, $f_z=200$ Hz and $f_r=2000$ Hz with $N=5\times10^6$.}
     \label{fig:li7_4fig}
\end{figure}

\begin{figure}[htbp]
     \includegraphics[width=3.2in]{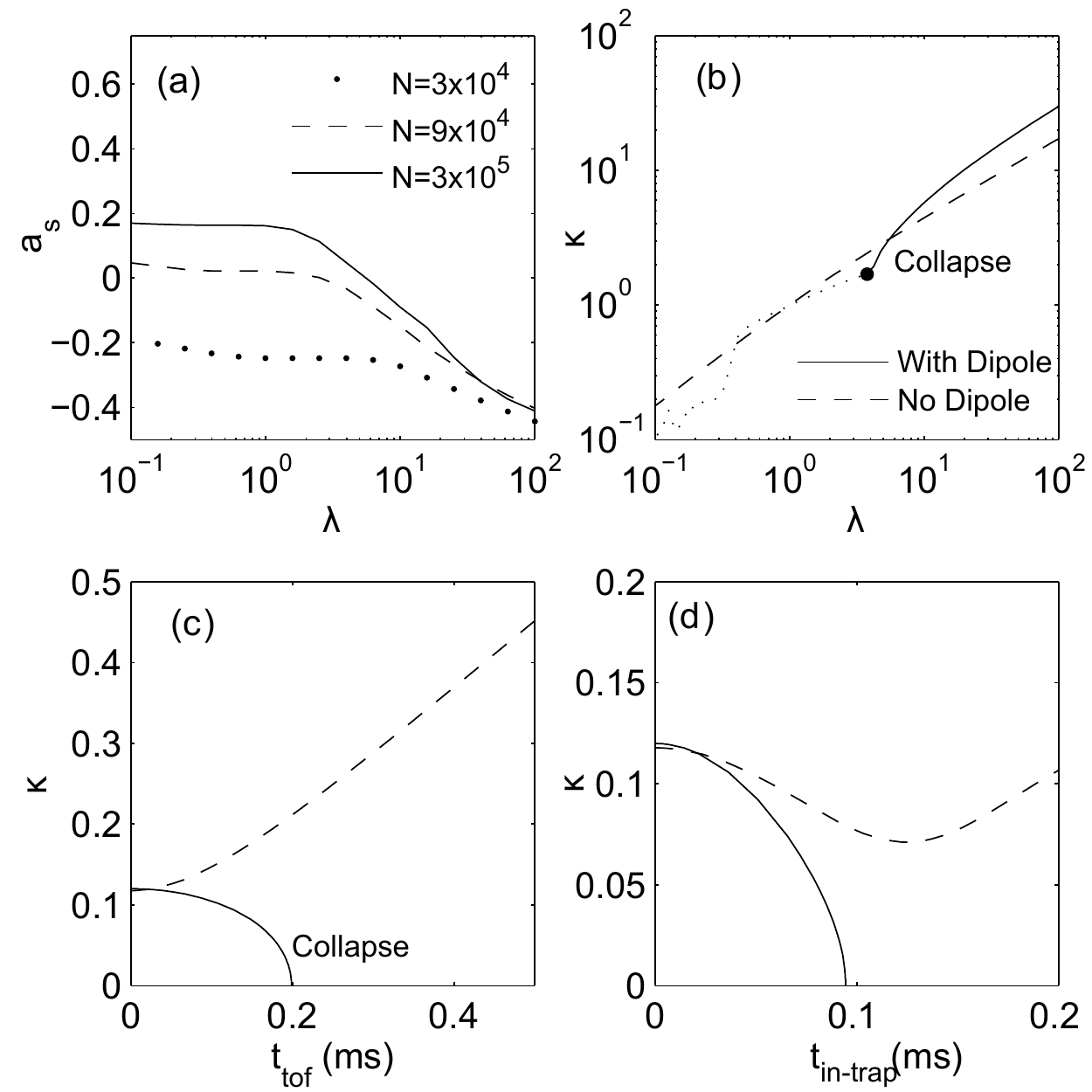}
	 \caption{Effect of MDDI in a $^{39}$K BEC. \textbf{(a)} the calculated threshold $a_s$ between stable and unstable regimes for a range of $\lambda$ and three representative atom numbers. \textbf{(b)} the BEC aspect ratio over a range of $\lambda$, with the collapse regime indicated by the faint dotted line, for which there is no stable solution found for Eqns. \ref{eqn:diffeqsRadial} and \ref{eqn:diffeqsAxial}. \textbf{(c)} Evolution of the aspect ratio of the BEC in TOF expansion after release from a trap. In the trap $a_s = 0.5 a_0$ (where the BEC is stable), and upon release $a_s$ is tuned to  $0.05 a_0$ to induce collapse in free expansion. \textbf{(d)} the in-trap evolution of a BEC similar to \textbf{(c)} with $a_s$ tuned to $0.05 a_0$ at $t=0$ to induce in-trap dipolar collapse. Parameters used in the simulation: \textbf{(a)} and \textbf{(b)}, $f_{avg}=700$ Hz; \textbf{(b)}, $N=5\times10^5$ and $a_s = 0.05 a_0$; \textbf{(c)} and \textbf{(d)}, $f_z=200$ Hz and $f_r=2000$ Hz with $N=5\times10^5$. }
     \label{fig:k394fig}
\end{figure}

\begin{figure}[htbp]
     \includegraphics[width=3.2in]{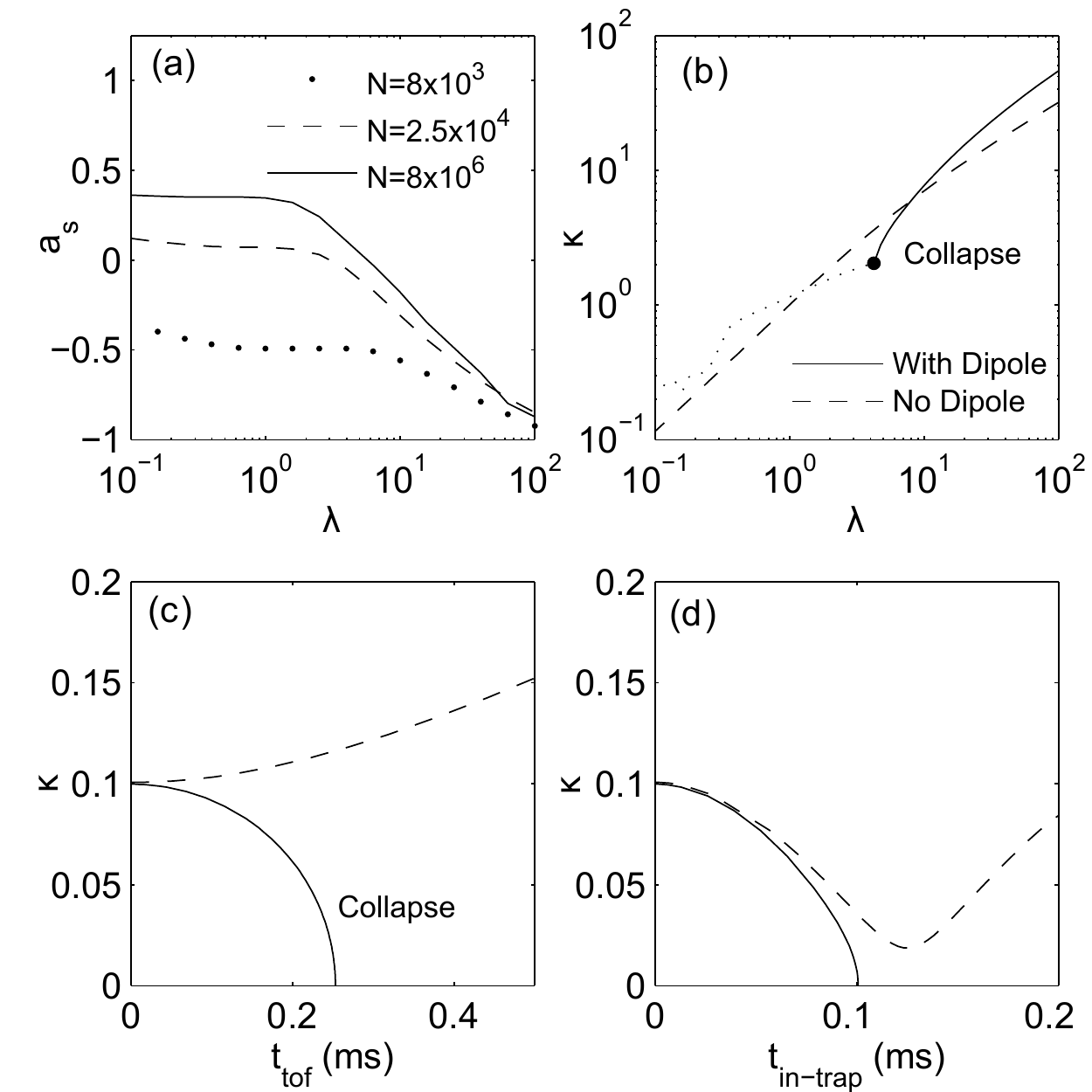}
	 \caption{Effect of MDDI in a $^{133}$Cs BEC. \textbf{(a)} the calculated threshold $a_s$ between stable and unstable regimes for a range of $\lambda$ and three representative atom numbers. \textbf{(b)} the BEC aspect ratio over a range of $\lambda$, with the collapse regime indicated by the faint dotted line, for which there is no stable solution found for Eqns. \ref{eqn:diffeqsRadial} and \ref{eqn:diffeqsAxial}. \textbf{(c)} Evolution of the aspect ratio of the BEC in TOF expansion after release from a trap. In the trap $a_s = 5 a_0$ (where the BEC is stable), and upon release $a_s$ is tuned to $0.1 a_0$ to induce collapse in free expansion. \textbf{(d)} the in trap evolution of a BEC similar to \textbf{(c)} with $a_s$ tuned to $0.1 a_0$ at $t=0$ to induce in-trap dipolar collapse. Parameters used in the simulation: \textbf{(a)} and \textbf{(b)}, $f_{avg}=700$ Hz; \textbf{(b)}, $N=2\times10^6$ and $a_s = 0.1 a_0$; \textbf{(c)} and \textbf{(d)}, $f_z=200$ Hz and $f_r=2000$ Hz with $N=2\times10^6$. }
     \label{fig:cs133_4fig}
\end{figure}

In  Fig.~\ref{fig:li7_4fig}, \textbf{(a)} shows the effect of $\lambda$ and $a_s$ on the stability of a BEC with MDDI. The line for each simulated BEC atom numbers $N$ indicates the boundary between the stable (above) and unstable (below) regimes. Similar to the $^{52}$Cr case (Fig.~\ref{fig:cr52stability}), the effect of MDDI is to stabilize a BEC in a pancake trap, and to destabilize a BEC in a cigar trap. \textbf{(b)} shows $\kappa$ over a range of $\lambda$, with the collapsed regime indicated by the dotted faint line. \textbf{(c)} shows plots of $\kappa$ after release from a trap. The in-trap starts with $a_s=0.01 a_0$ (in the stable regime), and upon release (at $t=0$ in the simulation) $a_s$ is tuned to $0.001 a_0$ to induce collapse in free expansion. Such rapid modulation of $a_s$ has already been performed experimentally in $^7$Li BECs \cite{Pollack_2010_arXiv}, and here could be used to reveal MDDI effects. \textbf{(d)} shows the in-trap evolution of a BEC initially at $a_s=0.01 a_0$ and then tuned to $0.001 a_0$ at $t=0$ to induce in-trap collapse due to MDDI. Neglecting MDDI would result in a stable, oscillating BEC, as seen in Fig.~\ref{fig:li7_4fig} \textbf{(d)}. The results in Fig.~\ref{fig:li7_4fig} offer four methods for detecting MDDI in $^7$Li BECs.  Similar calculations are provided for $^{39}$K and $^{133}$Cs, as seen in Fig.~\ref{fig:k394fig} and \ref{fig:cs133_4fig} respectively. We have shown that the MDDI can have substantial impact on the shape and stability for each of these alkali BECs.

\subsubsection{Other Alkalis}
While some of the other alkalis have potential for observing the effects of MDDI, the effects are usually small. For example, in $^{23}$Na and $^{85}$Rb BECs with atom number and trapping frequencies similar to current experiments (see Table \ref{tab:trapparameters}), a calculation including MDDI makes a five to ten percent difference in TOF aspect ratio from a calculation where MDDI are not included. Thus, though small, the MDDI effect lies within the bounds of possible experimental observations. For $^{41}$K and $^{87}$Rb, however, the Feshbach resonance is too narrow to provide the precision control of $a_s$ to carry out the type of experiments proposed here, and the MDDI effect is less than a one percent perturbation on the aspect ratio.

\subsection{Results for $^{164}$Dy and $^{168}$Er}
\begin{figure}[htbp]
     \includegraphics[width=3.2in]{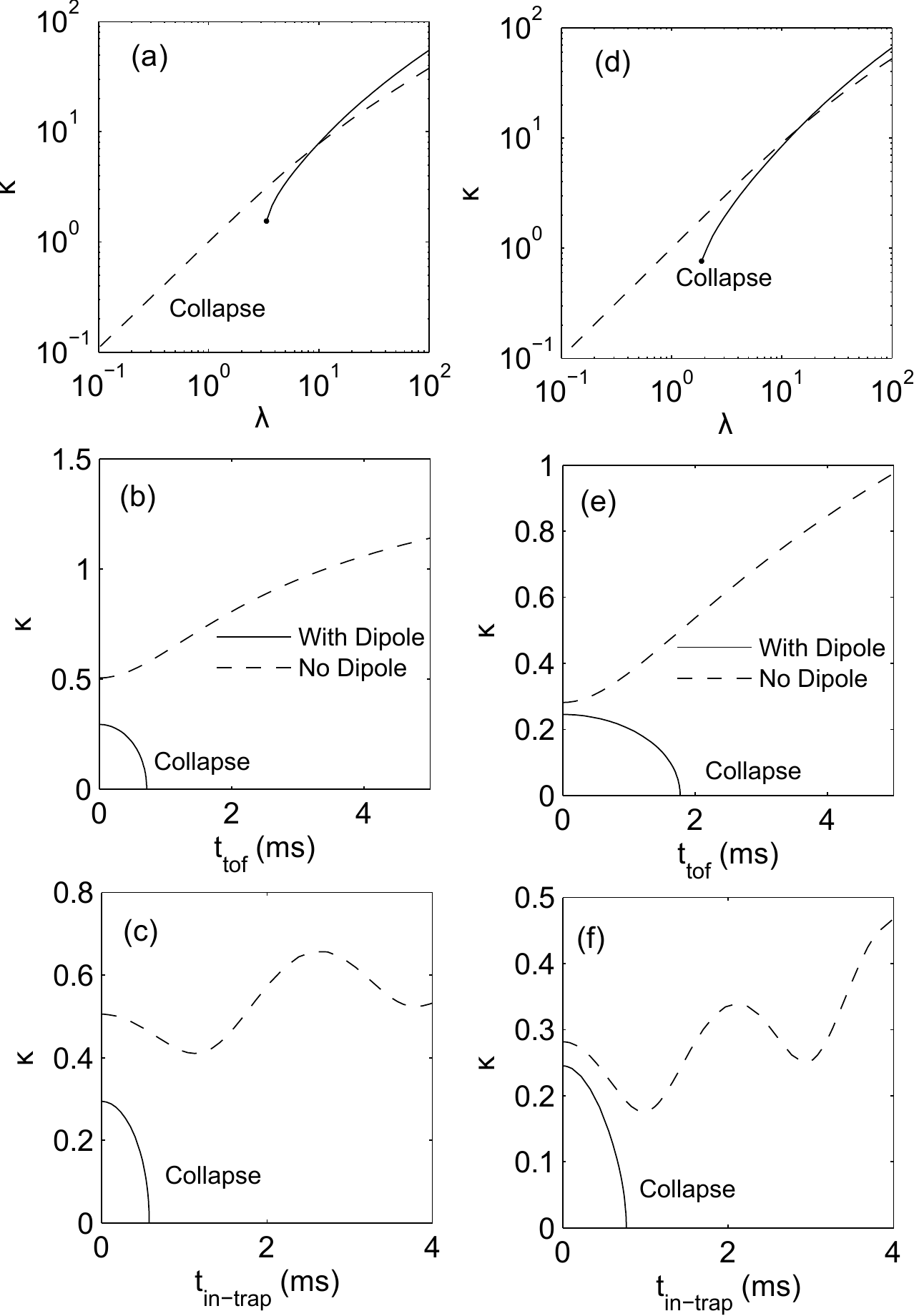}
	 \caption{Effects of MDDI in $^{164}$Dy and $^{168}$Er. The left column \textbf{(a-c)} contains results for $^{164}$Dy and the righ column \textbf{(a-c)} for $^{168}$Er. \textbf{(a,d)} the BEC aspect ratio over a range of $\lambda$, with the collapse regime indicated by the faint dotted line, for which there is no stable solution found for Eqns. \ref{eqn:diffeqsRadial} and \ref{eqn:diffeqsAxial}. \textbf{(b,e)} Evolution of the aspect ratio of the BEC in TOF expansion after release from a trap. In the trap $a_s = 150 a_0$ (where the BEC is stable for both species), and upon release $a_s$ is tuned to $75 a_0$ ($50 a_0$) for $^{164}$Dy ($^{168}$Er) to induce collapse in free expansion. \textbf{(c,f)} the in trap evolution of a BEC similar to \textbf{(b)} with $a_s$ tuned to $75 a_0$ ($50 a_0$) for $^{164}$Dy ($^{168}$Er) at $t=0$ to induce in-trap dipolar collapse. For all $^{164}$Dy calculations $N = 1.5 \times 10^4$, and for all $^{168}$Er $N = 7 \times 10^4$. For \textbf{(a,d)}, $f_{avg}=170$ Hz. For \textbf{(b,c)}, $f_z=100$ Hz and $f_r=200$ Hz, and for \textbf{(e,f)} $f_z=70$ Hz and $f_r=250$ Hz. }
     \label{fig:DyAndEr}
\end{figure}

Due to the much higher dipole moments in $^{164}$Dy and $^{168}$Er ($10 \mu_B$ \cite{Lu_PRL_2011} and $7 \mu_B$ \cite{McClelland_2006_PRL}, respectively) the effects of MDDI will be strongly apparent with larger values of $a_s$ than the alkalis. In Fig.~\ref{fig:DyAndEr} we present similar calculations to those of the alkali figures for these highly magnetic moment species, with the collapse dynamics occuring at higher values of $a_s$.

\section{Conclusion}
We showed that the variational method provides a useful and simple tool to simulate the effects of MDDI in BECs and presented various results for $^{52}$Cr and the alkalis.  For example, examining the the aspect ratio of freely expanding BECs should be sufficient for detecting the effects of MDDI in many species, and we suggest the investigation of $^{7}$Li, $^{39}$K , and $^{133}$Cs as favorable to detect such effects. We mention that future investigation of MDDI among non-alkali species looks promising as well. The achievement of BECs with $^{164}$Dy ($\mu=10\mu_B$)\cite{Lu_PRL_2011} and $^{168}$Er ($\mu=7\mu_B$) \cite{Aikawa_PRL_2012} is quite exciting, as the $a_{dd}$ for $^{168}$Er and $^{164}$Dy are 66.3 $a_0$ and 131.5$a_0$, much greater than even that of $^{52}$Cr. The ability to tune $a_s$ by Feshbach resonance could make these species unparalleled for the observation of strong MDDI effects. The method presented here is applicable to both species. To close, we highlight that the effects of MDDI on the BEC shape provide a clear and intuitive picture of MDDI in BECs. The examination of the BEC aspect ratio, for example, can be used as a sensitive measurement of the $a_s$ value in situ, and may prove a helpful calibration method for future studies of other---perhaps more exotic---MDDI effects. We also note that while our discussion is limited to magnetic dipole-dipole interactions in this paper, the variational method we present is general for all dipolar BECs and may be employed in calculating effects of electric dipole-dipole interactions (e.g. polar molecular BECs) as well \cite{Yi_2001_PRA}.

\appendix
\section{Supplemental Experimental Parameters}\label{append:experimentalParameters}
Typical atom numbers and trap frequencies currently employed in experimental studies of alkali BECs from the literature are listed below to show the experimental feasibility of the simulations provided here (Table \ref{tab:trapparameters}). 
\begingroup
\squeezetable
\begin{table}[htbp]
\begin{tabular}{cccccc}
\hline
\hline
Species & $f_r$ (Hz) & $f_z$ (Hz) & Num of Atoms & Ref\\
\hline
$^{7}$Li & 193 & 3 & $3\times 10^5$  & \cite{Pollack_2009_PRL}\\
$^{23}$Na & 1500 & 150 & $3\times 10^5$ & \cite{Stenger_1999_PRL}\\
$^{39}$K & 65 to 74 & 92 & $3\times 10^4$  & \cite{Roati_2007_PRL}\\
$^{41}$K & 325 & 15 & $3\times 10^5$  & \cite{Kishimoto_2009_PRA}\\
$^{85}$Rb & 17 & 6.8 & $1\times 10^5$  & \cite{Thompson_2005_PRL}\\
$^{87}$Rb & 930 & 11 & $3.6\times 10^6$  & \cite{Marte_2002_PRL}\\
$^{133}$Cs & 14 & 14 & $1.6\times 10^4$  & \cite{Weber_2003_Science}\\
$^{52}$Cr & 600 & 370 & $3\times 10^4$  & \cite{Lahaye_2007_Nature}\\
\hline
\end{tabular}
\caption{An example list of representative parameters for alkali BECs from the literature.}
\label{tab:trapparameters}
\end{table}
\endgroup

\section{Calculating the Magnetic Moment}\label{append:MagMoment}
A key parameter for the variational simulation is the value of the magnetic dipole moment, as $a_{dd} \propto \mu^2$. In low fields, the magnetic moment is found from the Zeeman effect, where
\begin{equation}
	\Delta E_{|F m_F>}=\mu_B g_F m_F B_z 
\end{equation}
In such a case, $\mu= \mu_B g_F m_F$. For higher fields, however, the Breit-Rabi formula is used for the ground states of alkali atoms \cite{Corney2006Atomic}: 
\begin{equation}
	E=\frac{-\Delta E_{hfs}}{2(2 I +1)} + g_I \mu_B m_F B  \pm \frac{\Delta E_{hfs}}{2}\left(1+\frac{2 m_F}{I+1/2} x + x^2\right)^{1/2}
\end{equation}
where $m_F=m_I \pm 1/2$, $\Delta E_{hfs} = A_{hfs} (I+1/2)$ is the hyperfine splitting, $x=\frac{g \mu_B B}{\Delta E_{hfs}}$, and $g=g_J - g_I$.

The magnetic dipole moment is simply the derivative of the energy with respect to the magnetic field, resulting in
\begin{equation}
	\mu (B) = g_I \mu_B m_F \pm \frac{\frac{2 m_F}{2 I+1}+x}{2\left(1+\frac{4 m_F}{2 I +1} x + x^2\right)^{1/2}} g \mu_B
\end{equation}
This is used to calculate the magnetic moment, $\mu_{cross}$, at $B$ where $a_s=0$ of all the alkali species (see Table \ref{tab:magneticmoment}).
\begingroup
\squeezetable
\begin{table*}[htbp]
\begin{tabular}{cccccccccc}
\hline
\hline
Species & I$_{nuc}$ & $A_{hfs}$ (MHz) & g$_I$ & $|F, m_F>$ & $B_{\infty}$(G) & $\Delta$ (G)& $a_{bg}$ ($a_0$) & Calculated $\mu_{cross}/\mu_B$ & Ref \\
\hline
$^{7}$Li & 3/2 &  401.752 & -0.001182 & $|1, +1>$  & 736.8 & -192.3 & -24.5 & 0.94 & \cite{Pollack_2009_PRL}\\
$^{23}$Na & 3/2 &  885.813 & -0.000805 & $|1, +1>$ & 907 & 1 & 63  & 0.91 &\cite{Stenger_1999_PRL} \\
$^{39}$K & 3/2 &  230.859 & -0.000142 & $|1, +1>$ & 403.4 & -52 & -33 & 0.95& \cite{Roati_2007_PRL, DErrico_2007_NJP}\\
$^{41}$K & 3/2 &  127.007 & -0.000078 & $|1, -1>$ & 51.34 & 0.3 & 60.54  & 0.07  & \cite{Kishimoto_2009_PRA}\\
$^{85}$Rb & 5/2 &  1011.910 & -0.000294 & $|2, -2>$ & 155.4& 11 & -443 & -0.57  & \cite{Thompson_2005_PRL}\\
$^{87}$Rb & 3/2 &  3417.341 & -0.000995 & $|1, +1>$ & 1007.4 & 0.170 & 100.5 & 0.73 & \cite{Marte_2002_PRL, Volz_2003_PRA}\\
$^{133}$Cs & 7/2 &  2298.157 & -0.000399 & $|3, -3>$ & -11.7 & 28.7 & 1411.8  & -.75 & \cite{Weber_2003_Science, Chin_2004_PRA}\\
$^{52}$Cr & 0 & -- & -- & $|3, -3>$ & 589.1 & 1.4 & 112 & 6$^{\ast}$ & \cite{Werner_2005_PRL} \\
\hline
\end{tabular}
\caption{Various parameters for the atomic species discussed in this paper. Values for $g_I$ are from Ref.~\cite{Arimondo_1977_RMP}. Values for A$_{hfs}$ are from \cite{Corney2006Atomic}. We use $g_J=2.002 319 304 3622(15)$ from \cite{CODATA}. Other values are from references listed in the table and \cite{Kohler_2006_RMP}. $^{\ast}$ from Ref.~\cite{Werner_2005_PRL}.}
\label{tab:magneticmoment}
\end{table*}
\endgroup

\begin{acknowledgments}
	The authors thank R.M. Wilson, R.G. Hulet, Han Pu, Yi Su and Li You for discussions, and T. Lahaye and T. Koch for providing the $^{52}$Cr data. YPC acknowledges the support of the Miller Family Foundation and NSF Grant No. CCF-0829918. AJO was supported by the Department of Defense through the National Defense Science and Engineering Graduate (NDSEG) Fellowship Program. 
\end{acknowledgments}

\bibliography{textBib}

\end{document}